\documentclass[aps,prl,twocolumn,showpacs,epsfig,floats]{revtex4-1}
\usepackage{amsmath}
\usepackage{color}
\usepackage{graphicx}
\usepackage{epsfig}
\usepackage{amsfonts}
\usepackage{mathrsfs}
\usepackage{amsmath}
\usepackage{float}
\usepackage{physics}
\usepackage{verbatim}

\begin{document} 

\title{Chaos and quantum scars in Bose-Josephson junction coupled to a bosonic mode}
 
\author{Sudip Sinha and S. Sinha}
\affiliation{Indian Institute of Science Education and
Research-Kolkata, Mohanpur, Nadia-741246, India}
 
\date{\today}

\begin{abstract}
We consider a model describing Bose-Josephson junction (BJJ) coupled to a single bosonic mode exhibiting quantum phase transition (QPT). Onset of chaos above QPT is observed from semiclassical dynamics as well from spectral statistics. Based on entanglement entropy we analyze the ergodic behavior of eigenstates with increasing energy density which also reveals the influence of dynamical steady state known as $\pi$-mode on it. We identify the imprint of unstable $\pi$-oscillation as many body quantum scar (MBQS), which leads to the deviation from ergodicity and quantify the degree of scarring. Persistence of phase coherence in nonequilibrium dynamics of such initial state corresponding to the $\pi$-mode is an observable signature of MBQS which has relevance in experiments on BJJ. 
%%-------------

\end{abstract}

%\pacs{05.30.-d, 05.45.Mt}

\maketitle

{\it Introduction:}
Ergodicity in closed quantum system has attracted attention in recent years due to its implication in variety of interesting phenomena related to the nonequilibrium dynamics of quantum many body system \cite{K_Sengupta,Ehud}. The emergence of steady states in nonequilibrium dynamics of certain many body systems and its correspondence to the generalized Gibbs ensemble is an evidence of thermalization \cite{Rigol1,Langen1,Gring,Caux,Diptiman1,R_Fazio}. The eigenstate thermalization hypothesis (ETH) has been proposed to explain thermalization of such closed quantum systems at the level of individual eigenstates \cite{Deutsch,Srednicki,P_Reimann} and its connection with random matrix theory (RMT) has thoroughly been explored theoretically \cite{Izrailev1,Izrailev2,LF_Santos1,Torres2,Alessio}. On the other hand, there are certain many body systems which fail to thermalize and remain localized, giving rise to many body localization (MBL) \cite{MBL_review,Nandkishore,Altshuler,Torres} phenomena \cite{Bordia,Schreiber1,Alessio2,Abanin1,Moessener,Shlyapnikov,Sinha}. Both the extreme phenomena are related to the degree of ergodicity of closed quantum system, which is a subject of intense study. As an alternate route to thermalization and delocalization, underlying chaotic behaviour has also been investigated in certain many body systems \cite{Altland1,Sayak}. Because of easy tunability of parameters, the ultracold atomic system has become a testbed to study both the MBL and nonequilibrium many body dynamics related to ergodicity \cite{Schreiber1,MBL_expt1,MBL_expt2,MBL_expt3,MBL_expt4,quantum_cradle,
three_mode_BEC_chaos,Langen2,Meinert}.

Non-ergodic multifractal wavefunctions are intermediate between localized and extended states, which have been observed in disordered as well as other many body systems \cite{Review_MF,Stephan,Bogomonly2,LF_Santos2,Masudul}, giving rise to non-ergodic behaviour such as anomalous thermalization, non-ergodic to ergodic transition and deserves further investigation \cite{Kravtsov,Abanin,Luitz,SYK,Sinha_2,Dicke_OTOC}.

Apart from above mentioned phenomena, there are other means by which a quantum system can deviate from ergodicity and certain states may lead to the breakdown of ETH hypothesis. In a recent experiment on chain of ultracold Rydberg atoms \cite{Bernein}, the appearance of revival phenomena for certain specific initial states indicating the deviation from ergodicity has been attributed to the existence of many body quantum scars (MBQS) and its underlying mechanism has been analyzed theoretically in a series of recent works \cite{Turner,Lin,Chandran,Choi2,Schecter1}. After this experiment, MBQS has also been identified theoretically in other models \cite{Sanjay,Schecter2} as well in a recent experiment on dipolar gas \cite{dipolar_gas_expt}. Quantum {\it scar} in single particle wavefunction can be identified as reminiscence of unstable classical periodic orbits, which was first studied in the context of chaotic stadium \cite{Heller}. However in quantum many body system, the correspondence between scarred state and unstable classical orbits is not very obvious. Similarly, the connection between ergodicity in interacting quantum system and underlying chaotic dynamics remains unclear and deserves further investigation. Identifying such {\it scar} as reminiscence of unstable collective mode of many body system and to detect its imprint on ergodicity are main focus of the present study.

\begin{figure} 
    \centering
    \includegraphics[height=4.5cm,width=6cm]{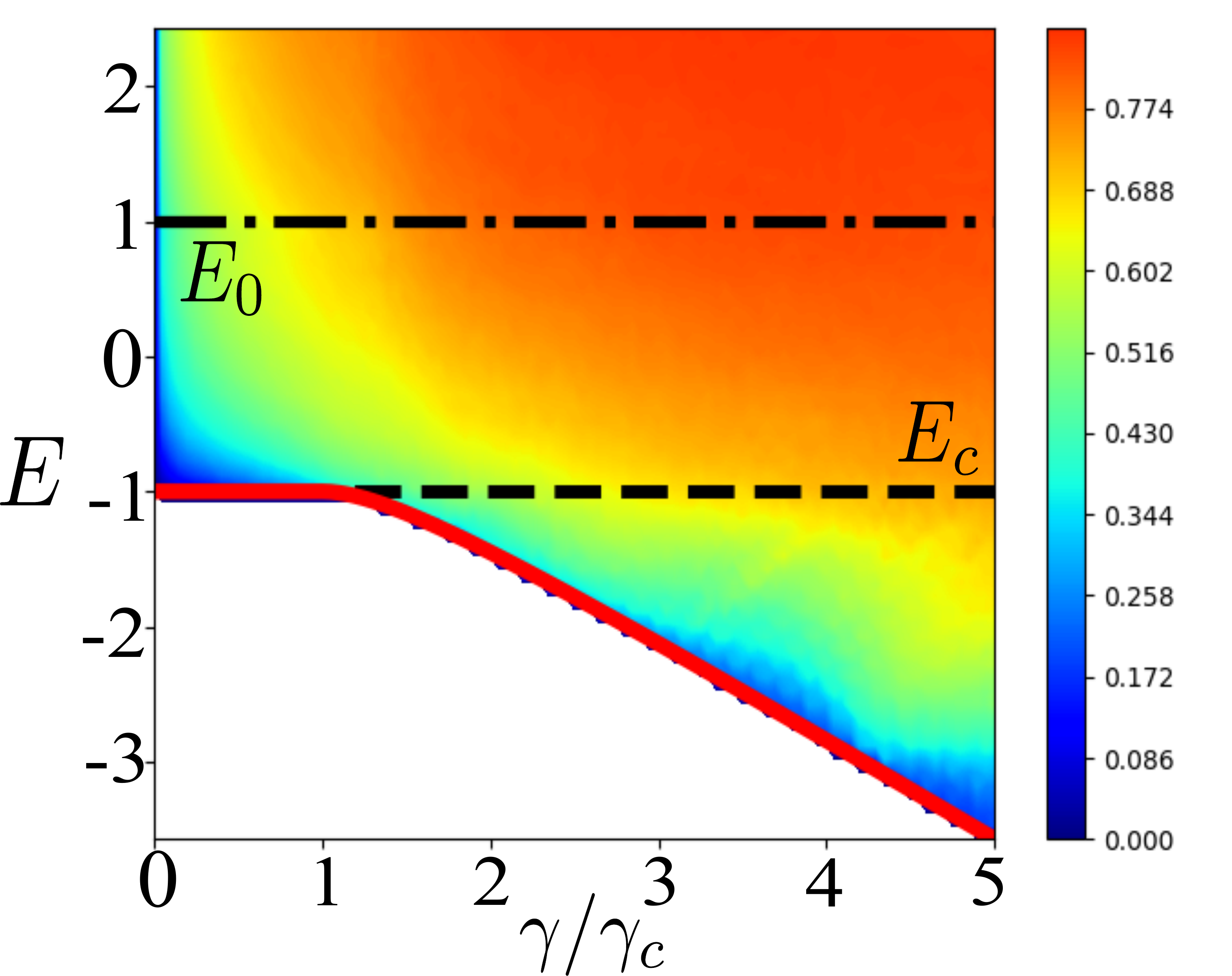} % scale=0.178
    \caption{Ergodic behaviour of BJJ based on relative entanglement entropy $S_{en}/S_{max}$ (in color scale) of eigenstates with energy density $E$ across the QPT ($\gamma/\gamma_c=1$). Different FPs: ground state (red solid line), unstable FP-II at $E_c$ (dashed line) and FP-III corresponding to $\pi$-mode at $E_0$ (dashed-dotted  line). Parameters chosen: $S=30$, $\omega_0=3$ and for all figures $U=0.5$, $N_{\text{max}} = 99$.}  
    \label{fig:1}
\end{figure}

In this work, we investigate the effect of dynamical steady states on ergodic behaviour and formation of quantum {\it scars} in a BJJ coupled to a single bosonic mode, which exhibits quantum phase transition (QPT) accompanied by onset of chaos. Similar connection between quantum {\it scar} in an interacting system and unstable dynamical modes of its classical counterpart, can also be explored in collective spin models \cite{coupled_top}. We quantify the ergodic behaviour of the states from entanglement entropy (EE), which is summarized in Fig.\ref{fig:1}. Interestingly, the presence of a dynamical steady state known as $\pi$-oscillation of BJJ \cite{Shenoy,Oberthaler,Pizzi} at an energy density $E_0$, gives rise to deviation from ergodicity and its imprint remains as {\it scar} in the wavefunction. Within single mode approximation, BJJ with fixed number of bosons $N$ can be described as two site Bose-Hubbard model (BHM) \cite{Walls}, $\hat{\mathcal{H}}_{BJJ}= -\frac{J}{2}(\hat{a}_{L}^{\dagger}\hat{a}_{R} + \hat{a}_{R}^{\dagger}\hat{a}_{L}) + \frac{U}{2N}\left[\hat{n}_{L}(\hat{n}_{L}-1) + \hat{n}_{R}(\hat{n}_{R} -1)\right]$, where $J$ and $U$ represent the hopping strength and on-site interaction respectively, while $\hat{a}_{L/R}$ and $\hat{n}_{L/R}$ denotes annihilation and number operators of boson in respective sites. The steady states and dynamics of BJJ has been studied extensively both experimentally \cite{Oberthaler,Levy,Albeiz} as well as theoretically \cite{Shenoy,Oberthaler,Pizzi,Walls,Polls,Vardi1,L_D_Carr,higher_modes,Minguzzi,
quantum_dynamics1,quantum_dynamics2,Vardi2,Kroha,Wimberger,Kastner,Sudip} 
%%%%%%%%%%%%%%%%%%%%% THE MODEL %%%%%%%%%%%%%%%%%%%%%%%%%%%%%%%%%

{\it Model.}--Using the Schwinger boson representation $\hat{S}_x=(\hat{a}^\dagger_{R}\hat{a}_{L}+\hat{a}^\dagger_{L}\hat{a}_{R})/2$ and $\hat{S}_z=(\hat{n}_{R}-\hat{n}_{L})/2$, $\hat{\mathcal{H}}_{BJJ}$ represents a large spin system with magnitude $S=
\frac{N}{2}$; as a result the BJJ coupled to a single bosonic mode can be described by the following spin Hamiltonian \cite{Sudip},
\begin{eqnarray}
\hat{\mathcal{H}}&=&-J\hat{S}_x+\frac{U}{2S}\hat{S}^2_z+\frac{\lambda\hat{S}_z}{2\sqrt{2S}}(\hat{b}+\hat{b}^\dagger)+\hbar\omega_0 \hat{b}^\dagger \hat{b} 
\label{spin_model}
\end{eqnarray}
where $\lambda$ is the coupling strength and $\hat{b}$ is annihilation operator of the bosonic mode with energy $\hbar \omega_0$. We redefine effective coupling strength as $\gamma=\lambda^2/\omega_0$ and scale energy (time) in units of $J$ ($1/J$) with $\hbar=1$. 

%%%%%%%%%%%%%%%%%%%%%%%%%% FIG HEAT MAP %%%%%%%%%%%%%%%%%%%%%%%%%%%%%%

%%%%%%%%%%%%%%%%%%%%%%%%%%%%%%%%%%%%%%%%%%%%%%%%%%%%%%%%%%%%%%%%%%%%%%

\begin{figure} 
    \centering
    \includegraphics[height=7.2cm,width=8.7cm]{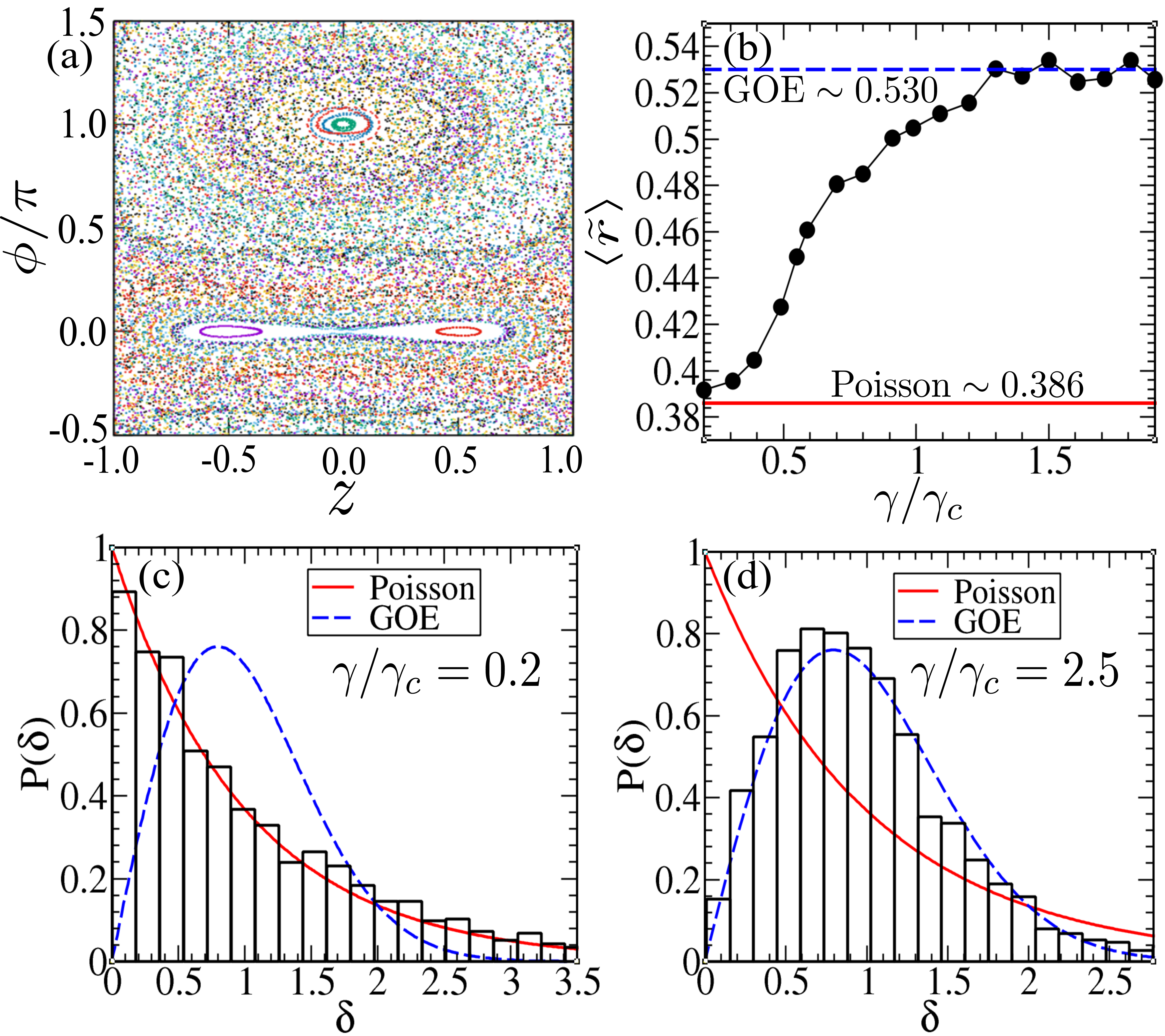}  %scale=0.115
    \caption{Onset of chaos: (a) Classical phase portrait for $\gamma/\gamma_c=1.11$, $\omega_0=3.0$ and $U=0.5$. (b) Crossover of $\langle \tilde{r} \rangle$ from Poissonian (red solid line) to GOE values (blue dashed line) by tuning $\gamma$. (c) and (d) Level spacing distributions for $\gamma<\gamma_c$ and $\gamma>\gamma_c$ respectively. Parameters chosen: $S = 25$ and for (b),(c),(d) $\omega_0=1.0$. For spectral statistics, first 1400 energy levels are considered from the even parity sector \cite{note_spectral}.}  
    \label{fig:2}
\end{figure}

%%%%%%%%%%%%%%%%%% SEMICLASSICAL ANALYSIS %%%%%%%%%%%%%%%%%%%%%%%%%
{\it Semiclassical analysis.}--
For $S=\frac{N}{2}\gg1$, the spin system is treated semiclassically by the wavefunction $\ket{\Psi_{sc}}=\ket{z,\phi} \otimes \ket{\alpha}$, where $\ket{z,\phi}$ and $\ket{\alpha}$ represent the coherent states of the spin and boson respectively \cite{Radcliffe}. The variables $\phi$ and $z=\cos{\theta}$ are canonical conjugate coordinates describing the orientation of the classical spin vector $\vec{S}=(S\sin\theta\cos\phi,S\sin\theta\sin\phi,S\cos\theta)$ while $\alpha=\sqrt{2S}(q+\imath p)/2$ denotes the dimensionless coordinates of the corresponding oscillator. In BJJ, $z$ and $\phi$ denote the number imbalance (magnetization) and relative phase between the two sites respectively \cite{Walls}. The corresponding classical Hamiltonian is given by \cite{Sudip}, 
\begin{eqnarray}
\mathcal{H}_{cl}=-\sqrt{1-z^2}\cos{\phi}+\frac{U}{2}z^2+\frac{\lambda}{2}zq+\frac{\omega_0}{2}(q^2+p^2)
\end{eqnarray}
The classical energy $E$ and $\mathcal{H}_{cl}$ are scaled by S to make them intensive. In isolated BJJ with $U>0$, the interaction is anti-ferromagnetic and the spin vector is aligned along x-axis giving $z=0$. Whereas in the presence of bosonic mode, the effective interaction $\tilde{U}=U-\gamma/4$ can become ferromagnetic, due to which BJJ undergoes a QPT at $\gamma_c=4(1+U)$ to a state having finite imbalance ($z\neq 0$) \cite{Sudip}. Such transition has also been confirmed from full quantum analysis \cite{Supplementary}.
The equations of motion (EOM) for collective coordinates can be obtained from Hamilton's equations \cite{Sudip,Supplementary},
\begin{equation}
\dot{\vec{X}}=\frac{\partial \mathcal{H}_{cl}}{\partial \vec{P}}, \,\,\,\,\,\,\, \quad \dot{\vec{P}}=-\frac{\partial \mathcal{H}_{cl}}{\partial \vec{X}}
\label{Hamilton}
\end{equation}
where $\vec{X}=\{\phi,q\}$ and $\vec{P}=\{z,p\}$ are canonical conjugates of each other.
The fixed points (FP) obtained from the EOM and their stability are analyzed near QPT \cite{Sudip}. For $\gamma>\gamma_c$, we categorize the FPs as: I. symmetry broken: $\{z=\pm\sqrt{1-(1/\tilde{U})^2},\phi=0,q=\mp\sqrt{\gamma z^2/4\omega_0},p=0\}$ with energy density $E_{GS}=-\frac{1}{2}(1/|\tilde{U}|+|\tilde{U}|)$, II. symmetry unbroken: $\{z=0,\phi=0,q=0,p=0\}$ with energy density $E_c=-1$ and III. corresponds to $\pi$-oscillation: $\{z=0,\phi=\pi,q=0,p=0\}$ \cite{Shenoy,Sudip} with energy density $E_0=1$. The stable FPs corresponding to I and III are visible in classical phase portrait (Fig.\ref{fig:2}(a)). Influence of this $\pi$-mode on ergodic behavior of the system is the main focus of present study.
%%%%%%%%%%%%%%%%%%%%%%%% CHAOS AND ERGODICITY %%%%%%%%%%%%%%%%%%%%%%

%%%%%%%%%%%%%%%%% FIG2 GROSS LEVEL CHAOS %%%%%%%%%%%%%%%%%%%%%%%%%%%%%
%%%%%%%%%%%%%%%%%%%%%%%%%%%%%%%%%%%%%%%%%%%%%%%%%%%%%%%%%%%%%%%%%%%%%%

{\it Ergodicity and chaos:}
Phase space trajectories corresponding to semiclassical dynamics of BJJ can be studied from Eq.\ref{Hamilton} \cite{Sudip}. In Fig.\ref{fig:2}(a), the phase portrait reveals chaotic dynamics near QPT indicating ergodic behavior. To analyze the system quantum mechanically, we obtain eigenvalues $\mathcal{E}_n$ and eigenfunctions $|\psi_n\rangle=\sum^{\mathcal{N}}_{i}\psi^{i}_{n}\ket{i}$ of the Hamiltonian (Eq.\ref{spin_model}) using the basis states $\ket{i}=\ket{m_z,n}$, where $m_z$ and $n$ are eigenvalues of $\hat{S}_z$ and number operator for bosonic mode respectively, where the latter is truncated to $N_{max}$ for numerical calculation. The dimension of the associated Hilbert space is $\mathcal{N}=(2S+1)(N_{max}+1)$. The eigenvectors and corresponding eigenvalues can be divided into even and odd sector of parity operator $\hat{\Pi}=e^{\imath \pi \hat{\mathcal{P}}}$, where $\hat{\mathcal{P}}=\hat{n}-\hat{S}_x+S$. The occurrence of excited state quantum phase transition (ESQPT) at FP-II leads to the suppression of energy gap between the consecutive even and odd parity states below $E_c$ separating the symmetry broken and unbroken states \cite{Supplementary,Caprio,chaos_dicke_ESQPT,Cejnar,Brandes,LF_Santos3}. To investigate the signature of chaos at the quantum level, we study the spectral properties. The eigenvalues $\mathcal{E}_{n}$ of each parity sector are sorted in ascending order and  spacing distribution P$(\delta)$ of the level spacing $\delta_n = \mathcal{E}_{n+1} - \mathcal{E}_{n}$ is obtained following the usual prescription \cite{Haake,Mehta}. According to Bohigas-Giannoni-Schmit (BGS) conjecture \cite{BGS}, the level spacing distribution of classically chaotic system follows Wigner-Dyson (WD) statistics, whereas Poissonian statistics, P$(\delta)=\exp\left(-\delta\right)$ can be observed in regular (integrable) regime \cite{Berry,Haake}. For weak coupling strength $\gamma < \gamma_c$, the level spacing follows Poisson distribution as evident from Fig.\ref{fig:2}(c). On the other hand, for increasing $\gamma$ above $\gamma_c$, the level spacing distribution resembles WD statistics, P$(\delta)=\frac{\pi\delta}{2}\exp(-\pi\delta^2/4)$ corresponding to Gaussian orthogonal ensemble (GOE) of RMT (see Fig.\ref{fig:2}(d)). Onset of chaos is also evident from the average value of level spacing ratio defined as $\langle \tilde{r} \rangle =\langle \text{min}(\delta_n,\delta_{n+1})/\text{max}(\delta_{n},\delta_{n+1}) \rangle$, which shows a crossover from Poissonian limit with $\langle \tilde{r} \rangle \sim 0.386$ to GOE limit with $\langle \tilde{r} \rangle \sim 0.530$ \cite{r_avg_ref} across the QPT as seen in Fig.\ref{fig:2}(b). Such spectral analysis indicates chaotic behavior above QPT, however detailed study of eigenvectors can reveal more interesting phenomena related to ergodicity and nonequilibrium dynamics. 
%%%---------------------------------

\begin{figure}[H]
    \centering
    \includegraphics[height=3.9cm,width=8.8cm]{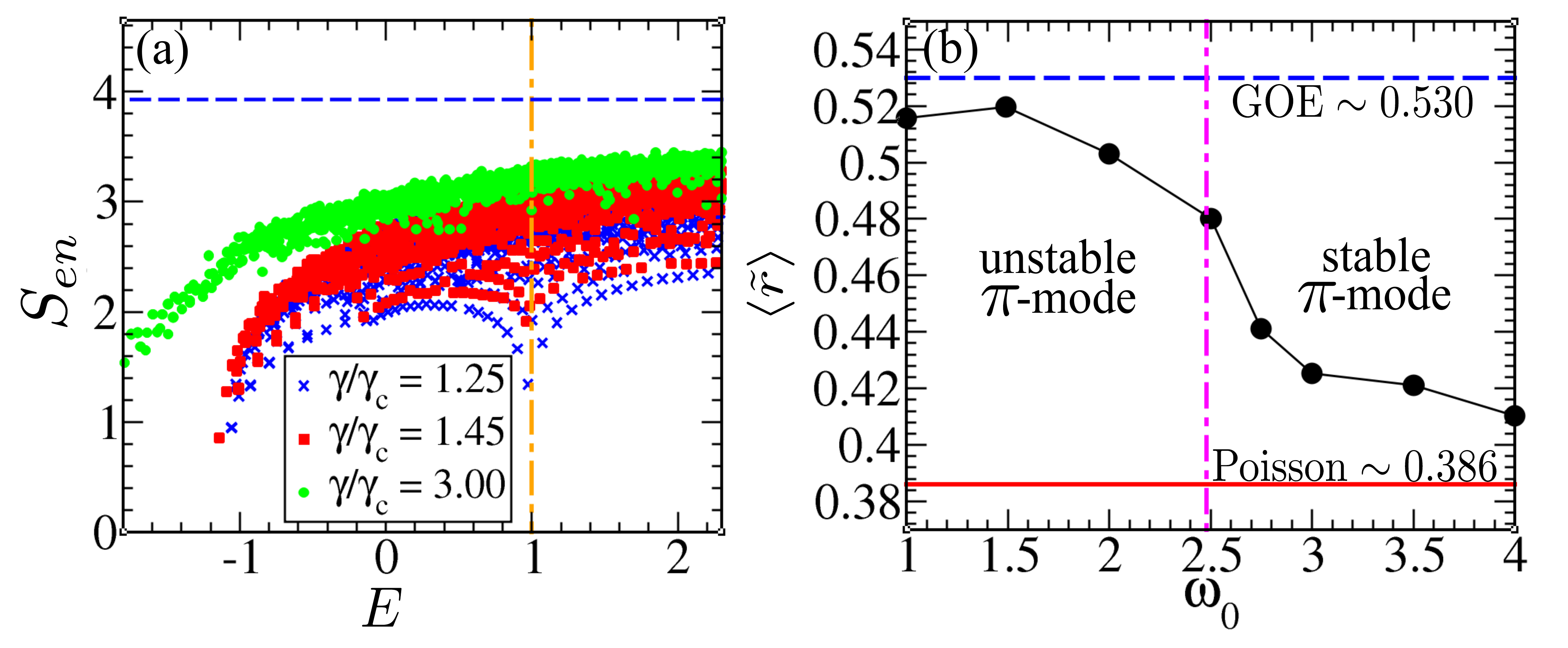}  %scale=0.115
    \caption{(a) EE of eigenstates with increasing energy density $E$ for different $\gamma/\gamma_c$ for $\omega_0=3$. Vertical dashed-dotted line: energy density $E_0$ of $\pi$-oscillation and horizontal  dashed line: $S_{max}$. (b) Variation of $\langle \tilde{r} \rangle$ with $\omega_0$ for fixed $\gamma/\gamma_c=1.2$. The pink dashed dotted line at $\omega_0=2.48$ divides the region of stability of $\pi$-mode. Parameters chosen: $S = 25$.} 
	\label{fig:3}
\end{figure}

%%%%%%%%%%%%%%%%%%%%%% EFFECT OF STEADY STATES %%%%%%%%%%%%%%%%%%%
{\it Effect of steady states on ergodic behavior:}
 In order to quantify the degree of ergodicity, we study the EE of energy eigenstates, which is also relevant for understanding non equilibrium properties of the BJJ. By tracing out the bosonic degrees, we obtain the reduced density matrix of the spin sector, $\hat{\rho}_S = \text{Tr}_{B}|\psi\rangle \langle \psi|$ and corresponding EE, $S_{en} =-\text{Tr}(\hat{\rho}_S\text{log}\hat{\rho}_S)$. To quantify the ergodic nature of a state, we compare it with the EE of maximally random state partitioned into subsystems A(B) with dimensionality $\mathcal{D}_{A}$($\mathcal{D}_{B}$) \cite{Page}.
Maximum EE corresponding to smaller subsystem A with $\mathcal{D}_{A} \ll \mathcal{D}_{B}$ can be written as,
\begin{eqnarray}
S_{max} \simeq \text{log}(\mathcal{D}_A)-O(\mathcal{D}_A/2\mathcal{D}_B)
\label{ent_entropy}
\end{eqnarray} 
For the present system $\mathcal{D}_{A} = 2S +1$ , $\mathcal{D}_{B}=N_{max} +1$ represent dimensions of spin and bosonic sector respectively. Relative EE, $S_{en}/S_{max}$ is obtained for eigenstates with increasing energy density $E$ and varying the coupling $\gamma$, which is shown as color scale plot in Fig.\ref{fig:1}, summarizing the ergodic behavior of the BJJ across QPT. It is evident from Fig.\ref{fig:1} and Fig.\ref{fig:3}(a), above QPT, the EE of states increases monotonically with energy density $E$ and approaches to the maximum limit (given by Eq.\ref{ent_entropy}), suggesting a crossover from non-ergodic to weak ergodic behavior which can also be confirmed from the nonequilibrium dynamics of BJJ \cite{Supplementary}. A recent study shows such relative EE of a subsystem below its maximum limit can indicate non-ergodic (multifractal) nature of states \cite{EE_MF_relation}. Similar behavior in degree of chaos has also been observed in Dicke model \cite{chaos_dicke_ESQPT,Dicke_chaos}.

Interestingly in the weakly ergodic regime, presence of $\pi$-mode at an energy density $E_0 =1$ (FP-III) can influence the ergodic properties. Stable $\pi$-oscillation can exist above QPT and becomes dynamically unstable at a coupling strength,
\begin{eqnarray}
\gamma_{u}=(\omega^2_0+U-1)^2/\omega^2_0. \label{unstable_pi}
\end{eqnarray}
As seen from Fig.\ref{fig:3}(a)), a dip in EE appears at $E_0$ due to the presence of such stable $\pi$-oscillation which gradually vanishes above $\gamma_u$. By increasing the frequency of the bosonic mode $\omega_0$, enhanced stability of the $\pi$-mode and suppression of chaos is observed around the energy density $E=E_0$ \cite{Supplementary}, which is also reflected from the decrease in $\langle \tilde{r} \rangle$ with $\omega_0$, as the unstable $\pi$-mode becomes stable (see Fig.\ref{fig:3}(b)) \cite{note_0}.
 
%%%%%%%%%%%%%%%%% FIG3 EE and SUPPRESSION OF CHAOS %%%%%%%%%%%%%%%%%%%%%%%%%%%%%
%%%%%%%%%%%%%%%%%%%%%%%%%%%%%%%%%%%%%%%%%%%%%%%%%%%%%%%%%%%%%%%%%%%%%%

\begin{figure}[H]
    \centering
    \includegraphics[height=10.5cm,width=9.095cm]{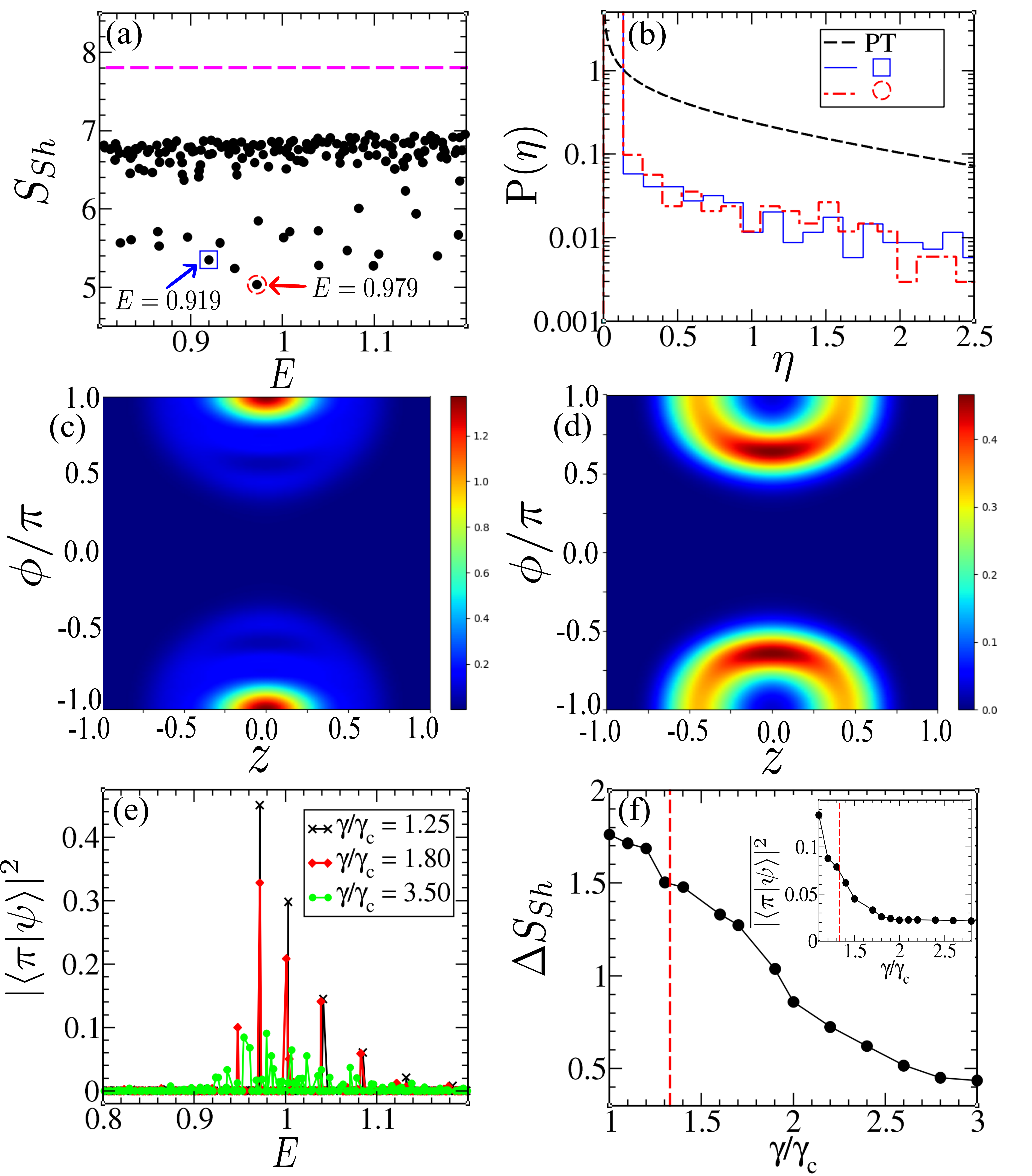} % scale=0.125
    \caption{MBQS of $\pi$-mode at $E_0$: (a) Shannon entropy in a small window near $E\approx E_0$. Horizontal dashed line represents the GOE limit. (b) Distribution $\text{P}(\eta)$ of scarred states (marked in (a)) at $\gamma = 1.8\gamma_c$. Husimi distribution $Q(z,\phi)$ of scarred states marked by (c) red circle and (d) blue square in (a). (e) Overlap of eigenstates with $\pi$-mode. Degree of scarring: (f) variation of $\Delta S_{Sh}$ ($\overline{|\bra{\pi}\ket{\psi}|^2}$ in the inset) with increasing $\gamma/\gamma_c$. Vertical red line denotes instability of $\pi$-mode at $\gamma_u=1.33\gamma_c$. Parameters chosen: $S=25$ and $\omega_0=3.0$.}
    \label{fig:4}
\end{figure}

{\it Quantum scar of $\pi$-oscillation:}
To search for imprint of unstable $\pi$-mode above the coupling strength $\gamma_u$ we closely analyze the eigenstates $\ket{\psi_n}$ within a small window of energy density around $E_0$. Unlike EE, Shannon entropy (SE) of states, $S_{Sh}=-\sum_{i}|\psi^{i}_{n}|^2\text{log}|\psi^{i}_{n}|^2$ within this window reveals an interesting structure as shown in Fig.\ref{fig:4}(a). The SE of most of the states form a band like structure below its GOE limit $\text{log}(0.48\mathcal{N})$ \cite{Izrailev1,Izrailev2}, whereas a few states with lower SE are deviated from this band. We identify scar of unstable $\pi$-oscillation in these deviated states, which become evident from the Husimi distribution $Q(z,\phi)=\frac{1}{\pi}\bra{z,\phi}\hat{\rho}_S\ket{z,\phi}$ of reduced density matrix $\hat{\rho}_S$ as depicted in Fig.\ref{fig:4}(c),(d). We also find that certain eigenstates around energy density $E_0$ have large overlap with the coherent state $\ket{\pi}=|z=0,\phi=\pi\rangle \otimes |q=0,p=0\rangle$, describing the steady state FP-III of $\pi$-mode.
The state with maximum overlap with $\ket{\pi}$ state turns out to be the most deviated state (marked by circle in Fig.\ref{fig:4}(a)) due to the scar of $\pi$-mode which is visible from its Husimi distribution with maximum phase space density at $\{z=0,\phi=\pi\}$ (see Fig.\ref{fig:4}(c)). More interestingly, the Husimi distribution of another deviated state (marked by square in Fig.\ref{fig:4}(a)) exhibits scar of an orbit around FP-III, corresponding to unstable $\pi$-oscillation (see Fig.\ref{fig:4}(d)). The trajectories around steady state of $\pi$-mode are also unstable with positive Lyapunov exponent and scar of such unstable $\pi$-oscillations are observed below energy density $E_0$.
%-------------------------------------------------
Next, we analyze the statistical properties of the wavefunction of such scarred states and compute the distribution P($\eta$) of the scaled elements $\eta=|\psi^{i}_n|^2\mathcal{N}$, which significantly deviates from Porter-Thomas (PT) distribution $\text{P}(\eta)=(1/\sqrt{2\pi\eta})\exp(-\eta/2)$ \cite{Haake} of eigenvectors of GOE matrices (see Fig.4(b)). This indicates non-Gaussian distribution of elements $\psi^{i}_n$ of scarred wavefunctions leading to the violation of Berry's conjecture \cite{Berry_conjecture}.

%%%%%%%%%%%%%%%%%%%%% FIG4: SCARS %%%%%%%%%%%%%%%%%%%
%%%%%%%%%%%%%%%%%%%%%%%%%%%%%%%%%%%%%%%%%%%%%%%%%%%%%%%%%%%%%%%%%%%

Above $\gamma_u$, the instability exponent of $\pi$-mode increases as $\Im(\omega) \sim \sqrt{\gamma-\gamma_u}$, which reduces the degree of scarring (DOS) resulting in enhancement of ergodicity. We quantify the DOS from the average deviation $\Delta S_{Sh}$ of SE of the scarred states from the band of weakly ergodic states. Both $\Delta S_{Sh}$ and the average overlap $\overline{|\bra{\pi}\ket{\psi}|^2}$ decreases with increasing coupling $\gamma$, as shown in Fig.\ref{fig:4}(f), indicating reduction of DOS. As a result of enhanced ergodicity, the corresponding distribution $\text{P}(\eta)$ also approaches to GOE limit with increasing $\gamma$.

%----------------------------------------------------
%%%%%%%%%%%%%%%%%%% FIG5 : DYNAMICS OF SCARRED STATES %%%%%%%%%%%%%%%%%%%%%%%%
\begin{figure}[H]
    \centering
    \includegraphics[height=6.955cm,width=9.075cm]{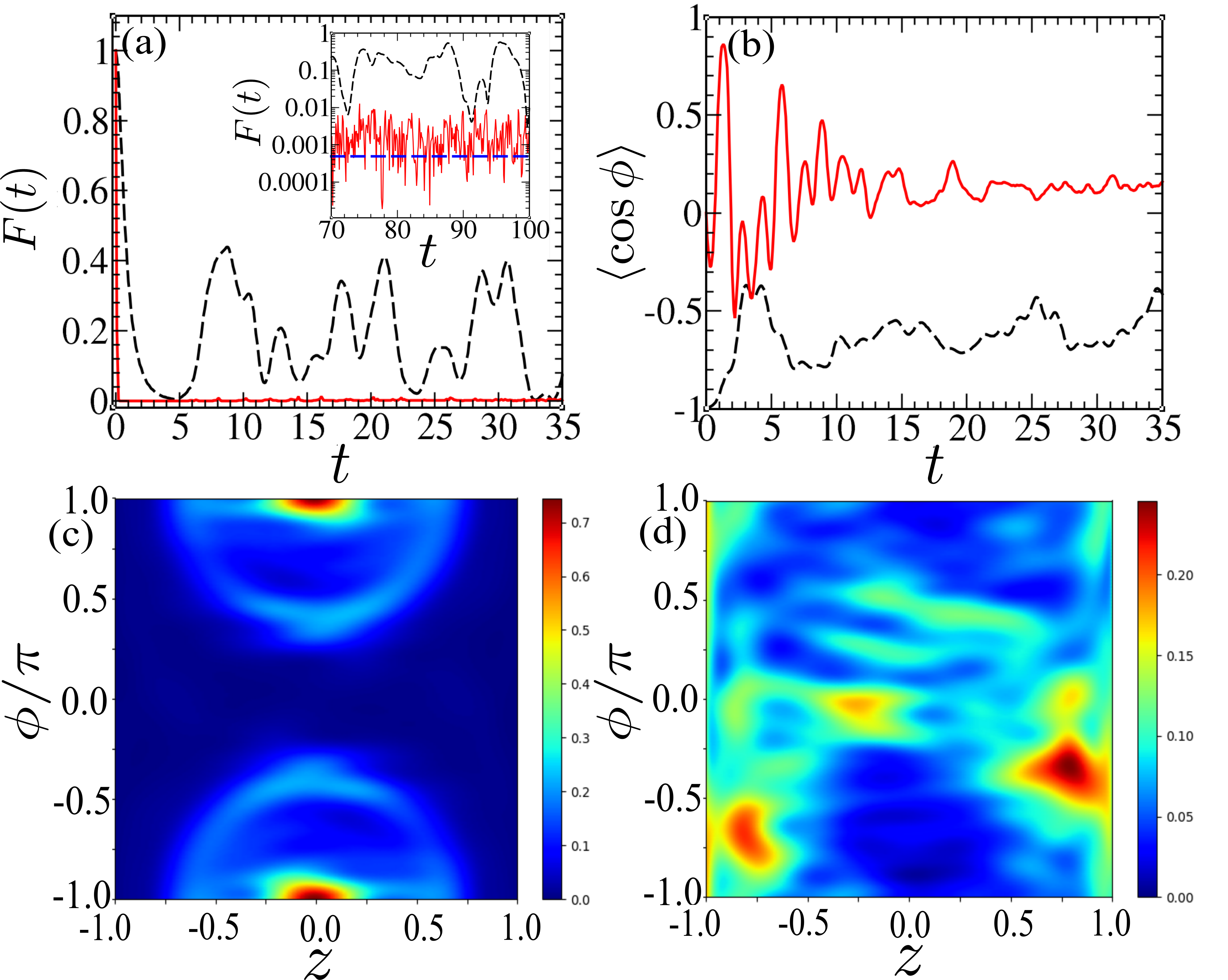} % scale=0.18
    \caption{Time evolution of $\ket{\pi}$ state (black dashed curve) and arbitrary state $|z=0.60,\phi=\pi/2\rangle \otimes |q=0.40,p=0.0\rangle $ (red solid curve) at $E\approx E_0$ shown by (a) survival probability $F(t)=|\bra{\psi(t)}\ket{\psi(0)}|^2$ and (b) phase coherence. Inset in (a) shows long time behaviour of $F(t)$ compared with GOE limit. (c), (d) show Husimi distribution of $\ket{\pi}$ state and arbitrary state respectively after sufficient time $t = 25$. Parameters chosen: $S=30$, $\gamma=1.8\gamma_c>\gamma_u$, and $\omega_0=3.0$.}
    \label{fig:5}
\end{figure}
%%%%%%%%%%%%%%%%%%%%%%%%%%%%%%%%%%%%%%%%%%%%%%%%%%%%%%%%%%%%%%%%%%%%%%%%  

Effect of such quantum scar can also be observed from anomalous behavior of nonequilibrium dynamics. Time evolution of the $\ket{\pi}$ state exhibits interesting features compared to any other arbitrary state at similar energy density (see Fig.\ref{fig:5}). Survival probability $F(t)=|\bra{\psi(t)}\ket{\psi(0)}|^2$ of this initial state shows oscillatory behavior and in long time deviates significantly from the GOE limit $3/\mathcal{N}$ \cite{Torres2,Izrailev2} compared to other initial states (see Fig.\ref{fig:5}(a)). Also the Husimi distribution of the final state in the time evolution retains the scar of $\ket{\pi}$ state (see Fig.\ref{fig:5}(c)). From the phase operator $\hat{\phi}$ \cite{Supplementary,Gati} defining the relative phase between two sites of BJJ, we calculate $\langle \cos{\phi} \rangle$  quantifying the phase coherence \cite{Gati,Stringari,phase_expt}. Remarkably, such phase coherence persists during the time evolution of $\ket{\pi}$ state in contrast to its decay in ergodic dynamics (see Fig.\ref{fig:5}(b)), which can serve as an experimentally observable effect of quantum scar. 

{\it Conclusion.}--BJJ coupled to a single bosonic mode constitutes an interesting model exhibiting QPT, onset of chaos and athermal behaviour related to quantum {\it scars}. 
We identify quantum {\it scars} related to collective $\pi$-oscillation
and quantify degree of scarring. In contrast to ergodic dynamics, phase coherence is retained during time evolution for special choice of initial state corresponding to $\pi$-mode, which is a detectable signature of MBQS relevant for experiments on BJJ \cite{phase_expt}. This model can also be realized in experiment by coupling a BJJ with cavity mode similar to the experiments in \cite{Baumann,cavity_QED_raman} and also in circuit QED setup \cite{circuit_QED}. In absence of interaction ($U=0$), the present model reduces to Dicke model \cite{dicke_original}, which is also a promising candidate to study the effect of quantum scarring, due to its experimental realization \cite{Baumann,cavity_QED_raman}. 

Present study elucidates the formation of {\it scar} in an interacting quantum system and its connection with underlying collective dynamics, which can also be explored in other systems \cite{coupled_top}.\\

We thank Hans Kroha for discussion.


\begin{thebibliography}{99}
\bibitem{note_spectral}
For spectral statistics, we have taken energy levels up to a sufficiently high energy density, where the degree of chaos is not changing as well the truncation effect of the Fock space is minimized.
\bibitem{note_0}
As evident from Fig.\ref{fig:3}(b), the onset of chaos occurs at a faster rate for $\omega_0 \sim 1$ (as in Fig.\ref{fig:2}(b)-(d)), for which the $\pi$-oscillation remains unstable above QPT and is favourable for detection of chaos from spectral statistics.
\end{thebibliography}
\end{document}

\begin{widetext}

\begin{center}
{\bf SUPPLEMENTARY MATERIAL}
\end{center}

\section{Quantum Phase Transition (QPT)}
From the semiclassical analysis, it is clear that the model (Eq.1 of the main text) describing a Bose-Josephson junction (BJJ) coupled to a bosonic mode undergoes a quantum phase transition at a critical coupling strength $\gamma_c=4(1+U)$ \cite{Sudip}. We confirm the existence of QPT using full quantum mechanical analysis by calculating relevant physical quantities from the ground state $\ket{\Psi_g}$. As seen from Fig.\ref{fig:1}, sharp change in ground state energy, magnetization $\langle \hat{S}_{z}\rangle/S$ (number imbalance), boson number $\langle \hat{n}\rangle$ and peak in entanglement entropy (EE) at $\gamma_c$ signify the QPT. 
It can be easily seen that the the results obtained for different quantities converge towards those obtained from the semi classical analysis for large values of spin $S$. 

%%%%%%%%%%%%%%%%%% FIG1 QUANTUM PHASE TRANSITION %%%%%%%%%%%%%%%%%%%%%
\begin{figure}[H]
    \centering
    \includegraphics[height=10cm,width=13cm]{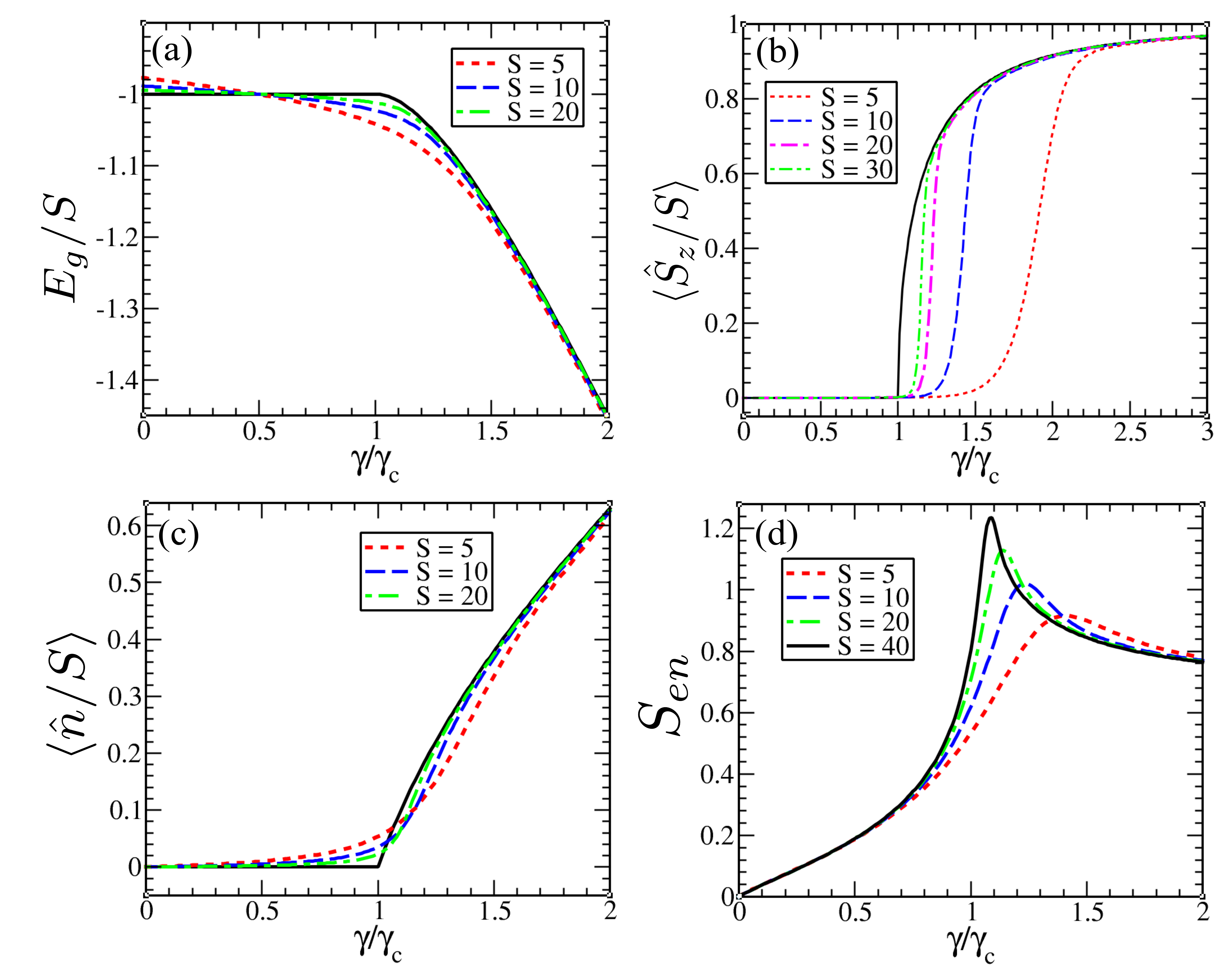}
    \caption{Quantum phase transition: (a) ground state energy, (b) magnetization (c) mean occupation number of bosons and (d) entanglement entropy as a function of coupling strength $\gamma$ for different values of spin $S$. solid curves in (a), (b) and (c) show the classical values. Parameters chosen for all figures are $\omega_0=2.0$ and $U=0.5$}
    \label{fig:1}
\end{figure}
%%%%%%%%%%%%%%%%%%%%%%%%%%%%%%%%%%%%%%%%%%%%%%%%%%%%%%%%%%%%%%%%%%%%%%

%%%%%%%%%%%%%%%%%%% DYNAMICAL EQUATIONS OF MOTION %%%%%%%%%%%%%%%%%%%%%%%%%%%%%
\noindent {\Large Dynamical equations of motion:}

The equations of motion obtained for collective coordinates $\{z,\phi,q,p\}$ \cite{Sudip} are given by,
\begin{subequations}
\begin{eqnarray}
&&\dot{z}=-\sqrt{1-z^2}\sin{\phi} \label{eqm1} \\
&&\dot{\phi}=\frac{z}{\sqrt{1-z^2}}\cos{\phi}+Uz+\sqrt{\gamma\omega_0}\frac{q}{2} \label{eqm2}\\
&&\dot{q}=p\omega_{0} \label{eqm3}\\
&&\dot{p}=-\sqrt{\gamma\omega_0}\frac{z}{2}-q\omega_{0} \label{eqm4}
\end{eqnarray}
\end{subequations}
From the above equations, we find that the stable steady state $\{z=0,\phi=0,q=0,p=0\}$ undergoes a pitchfork bifurcation at QPT, thus becoming unstable for $\gamma>\gamma_c$ and followed by formation of two new symmetry broken steady states which are represented by $\{z=\pm\sqrt{1-(1/\tilde{U})^2},\phi=0,q=\mp\sqrt{\gamma z^2/4\omega_0},p=0\}$ \cite{Sudip}.
Stability analysis of the steady state $\{z=0,\phi=\pi,q=0,p=0\}$ corresponding to the $\pi$-oscillation gives the frequency \cite{Sudip},
\begin{equation}
\omega^2 = \frac{1}{2}\bigg( \omega^2_0 + \omega^2_{su} \pm \sqrt{(\omega^2_0-\omega^2_{su})^2-\omega^2_0\gamma}\bigg) \quad \text{where} \quad \omega^2_{su}=1-U
\end{equation}

This frequency becomes imaginary at $\gamma_u$ (Eq.5 of the main text) indicating dynamic instability of $\pi$-oscillation and the instability denoted by the imaginary part $\Im(\omega)$ increases as $\sim\sqrt{\gamma-\gamma_u}$. 

\section{Excited State Quantum Phase Transition (ESQPT)}
Present model undergoes a excited state phase transition (ESQPT) above QPT ($\gamma>\gamma_c$) similar to the Dicke model \cite{Brandes}. The singularity at density of states signifies ESQPT at a critical energy density \cite{Brandes,Caprio,Cejnar,LF_Santos1}. In this case, the singularity in derivative of density of states signifies ESQPT at $E = E_c =-1$. This also corresponds to the energy of unstable symmetry unbroken steady state above QPT, which separates the symmetry broken states with energy $E<E_c$ from symmetry unbroken states for $E>E_c$. This can be captured from the pair gap between energy of consecutive odd and even parity states which is defined as  $\Delta = \mathcal{E}_{2n} - \mathcal{E}_{2n-1}$ where $n=1,2...$. The pair gap $\Delta$ is vanishingly small for symmetry broken sector with $E<E_c$, as shown in Fig.\ref{fig:2}(d). We also calculate other physical quantities depicted in Fig.\ref{fig:2} which confirm the ESQPT and separatrix between symmetry broken and unbroken states at $E_c$.

%%%%%%%%%%%%%%%%%%%%% FIGURE EXCITED STATE QPT%%%%%%%%%%%%%%%%%%%
\begin{figure}[H]
    \centering
    \includegraphics[height=10cm,width=12cm]{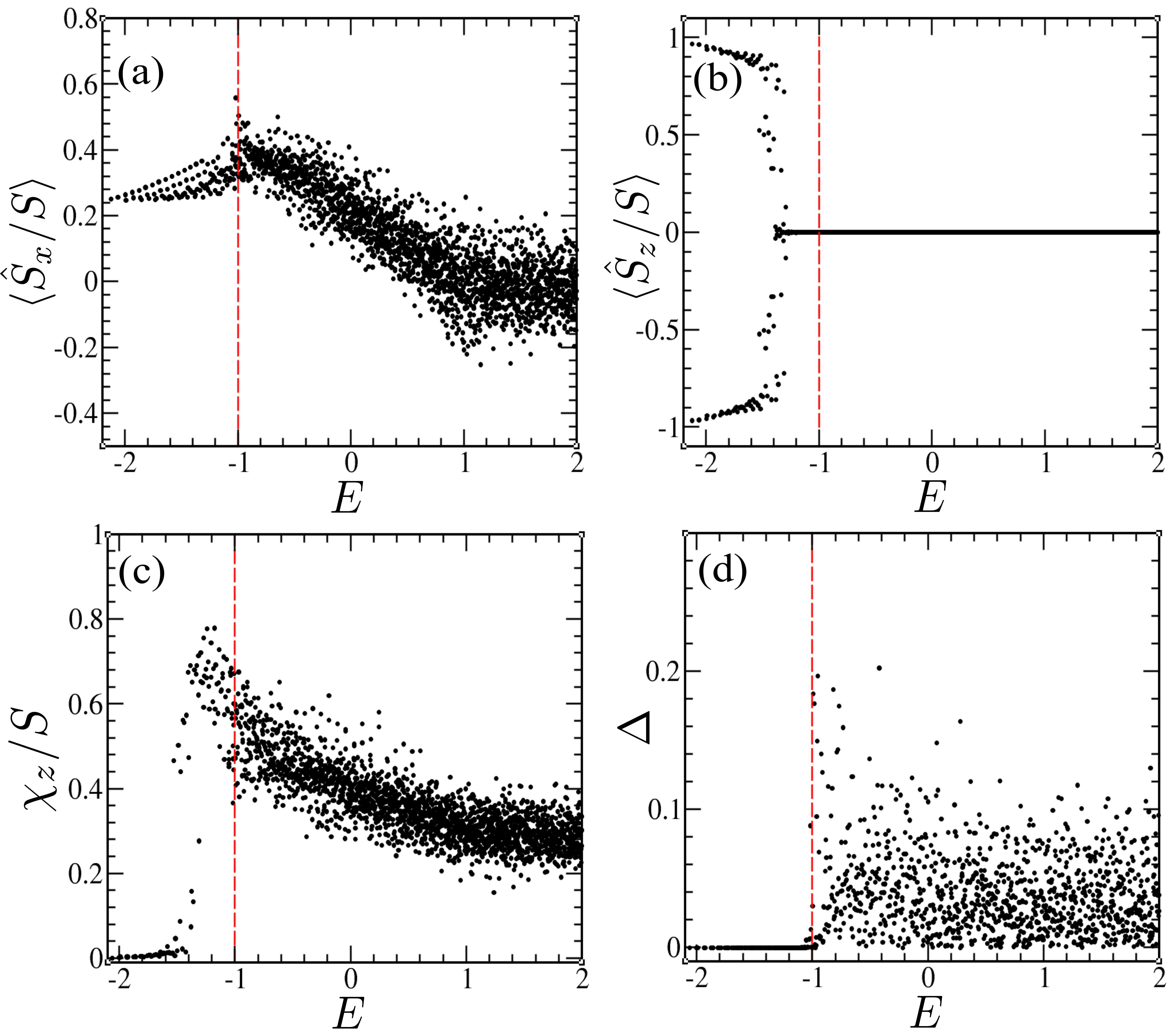}
    \caption{Signature of ESQPT captured by plotting (a) expectation of $\hat{S}_x/S$, (b) imbalance factor (magnetization) $\langle \hat{S}_z/S \rangle$, (c) fluctuations in $\hat{S}_z$  and (d) pair gap with scaled energy $E$ for $S=30$. Red dashed lines demarcates the ESQPT energy i.e $E=E_c$. Parameters chosen for all figures are $\gamma/\gamma_c=3.0$, $\omega_0=2.0$ and $U=0.5$. The red dashed line demarcates the energy at which ESQPT occurs i.e $E\approx E_c$.}
    \label{fig:2}
\end{figure}
%%%%%%%%%%%%%%%%%%%%%%%%%%%%%%%%%%%%%%%%%%%%%%%%%%%%%%%%%%%%%%%%%

\section{Effect of bosonic mode frequency}
For the model discussed in the main text, the onset of chaos occurs  by tuning the parameter $\gamma/\gamma_c$ above QPT. As a result, the spectral statistics exhibits a crossover across QPT from Poissonian to GOE limit, which is evident from the change in $\langle \tilde{r} \rangle$ with $\gamma/\gamma_c$, shown in Fig.2(b) of the main text. Although the overall chaotic behavior can be detected from spectral statistics, our studies show there is also an interesting effect of the bosonic mode frequency $\omega_0$. At the classical level, the stability of $\pi$-mode at $E=E_0$ can be controlled by tuning the mode frequency (given in Eq.5 of the main text). An increase in $\omega_0$ leads to enhanced stability of $\pi$-mode, which can exist as a stable fixed point in the chaotic regime above QPT for appropriate choice of $\omega_0$. To investigate the effect of such stable (unstable) $\pi$-mode, we study the Poincar\'e sections at fixed classical energy $E=E_0$, for different values of $\omega_0$, at a suitably chosen coupling strength $\gamma>\gamma_c$ in the chaotic regime (see Fig.\ref{fig:3}). The Poincar\'e sections are prepared by taking the intersection of the classical trajectories with the surface $q=0$ and projecting these points on the $\phi-z$ plane. At this fixed coupling strength ($\gamma_u<\gamma$) for $\omega_0=1$ , the $\pi$-oscillation become unstable and the Poincar\'e section shown in Fig.\ref{fig:3}(a) exhibits complete chaotic behavior. On the other hand, the $\pi$-mode can become the stable fixed point by tuning the mode frequency for larger value of $\omega_0=3$ ($\gamma_u>\gamma$), as a result mixed phase space behavior is observed along with regular orbits around $\phi=\pi$ (see Fig.\ref{fig:3}(b)), exhibiting the suppression of chaotic behavior at the classical energy $E=E_0$. This is also seen from the appearance of a dip in EE around the energy density $E=E_0$, when the $\pi$-mode remains stable for $\gamma<\gamma_u$, which slowly vanishes above $\gamma_u$ (see Fig.3(a) of the main text). To quantify this dip in EE, we compute the deviation of minimum EE, $\Delta S_{dip}$ from the average EE computed for eigenstates within a small window of energy density around $E=E_0$. For a fixed coupling strength $\gamma$ above the QPT, a significant increase of $\Delta S_{dip}$ is observed above the mode frequency for which the unstable $\pi$-mode becomes stable (see Fig.\ref{fig:3}(c)), indicating a reduction in degree of ergodicity around the said energy density. This behavior is also consistent with decrease in $\langle \tilde{r} \rangle$  with $\omega_0$ at this fixed value of $\gamma$, as the unstable $\pi$-mode becomes stable, depicted in Fig.\ref{fig:3}(b) of the main text.

%%%%%%%%%%%%%%%% EFFECT OF PI-MODE  %%%%%%%%%%%%%%%%%%%%
\begin{figure}[H]
\centering
\includegraphics[height=5.5cm,width=18.1cm]{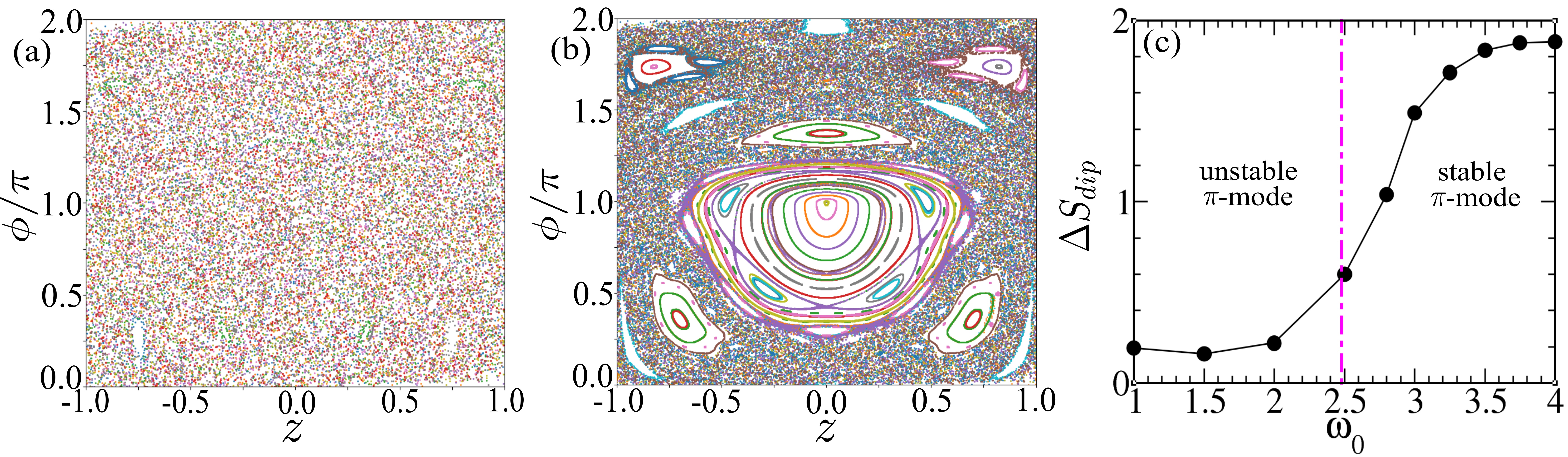}
\caption{a) Poincar\'e sections at classical energy $E=E_0$, $\gamma=1.2\gamma_c$, exhibiting (a) chaotic behavior for  $\omega_0=1$ ($\gamma_u<\gamma$) and (b) mixed phase space behavior for $\omega_0=3$ ($\gamma_u>\gamma$). (c) Variation of $\Delta S_{dip}$ with $\omega_0$ for $\gamma=1.2\gamma_c$. The pink dashed dotted line at $\omega_0=2.48$ divides the region of stability of $\pi$-mode. }
    \label{fig:3}
\end{figure}
%%%%%%%%%%%%%%%%%%%%%%%%%%%%%%%%%%%%%%%%%%%%%%%%%%%%%%%%%%%%%%%%%%%%%%

%%%%%%%%%%%%%%%%%%%%%%%%%%%%%%%%%%%%%%%%%%%%%%%%%%%%%%%%%%%%%%%%%%%%%
\section{Non equilibrium dynamics of non-ergodic vs ergodic states}
In this section, we show how the non-ergodic and weakly ergodic regime can be distinguished from non equilibrium dynamics. We compare various physical quantities obtained from time evolution of two initially chosen coherent states $\ket{z,\phi} \otimes \ket{\alpha}$
with energy $E<E_c$ and $E>E_c$ as shown in Fig.\ref{fig:4} and Fig.\ref{fig:5}.\\

For initial state with  $E<E_c$, the magnetization $\langle \hat{S}_z/S \rangle$ oscillates around a finite non-zero value which is indicative of the symmetry broken sector whereas for $E>E_c$, it decays to zero and the other components of spin coincide with it indicating thermalization (see Fig.\ref{fig:4}). We study the reduced density matrix corresponding to the spin sector $\hat{\rho}_S$ (in the $S_z$ basis) after sufficiently long time, starting from these two initial states shown in Fig.\ref{fig:5}. For $E<E_c$, non-ergodic dynamics is reflected in  non-diagonal and localized form of the final density matrix (Fig.\ref{fig:5}(b)). On the other hand for $E>E_c$, density matrix takes the diagonal form (Fig.\ref{fig:5}(c)) indicating thermalization to microcannonical ensemble, however certain fluctuations are observed due to weak ergodicity of the states. The time evolution of entanglement entropy (EE) in the ergodic regime approaches to $S_{max}$ (Eq.4 of the main text), whereas it deviates significantly for $E<E_c$, as shown in Fig.\ref{fig:5}. Similar distinction has also been observed from the long time behaviour of survival probability.\\

%%%%%%%%%%%%%%%%%%%% FIGURE EROGIC VS NON ERGODIC %%%%%%%%%%%%%%%%%%%%
\begin{figure}[H]
    \centering
    \includegraphics[height=6cm,width=14cm]{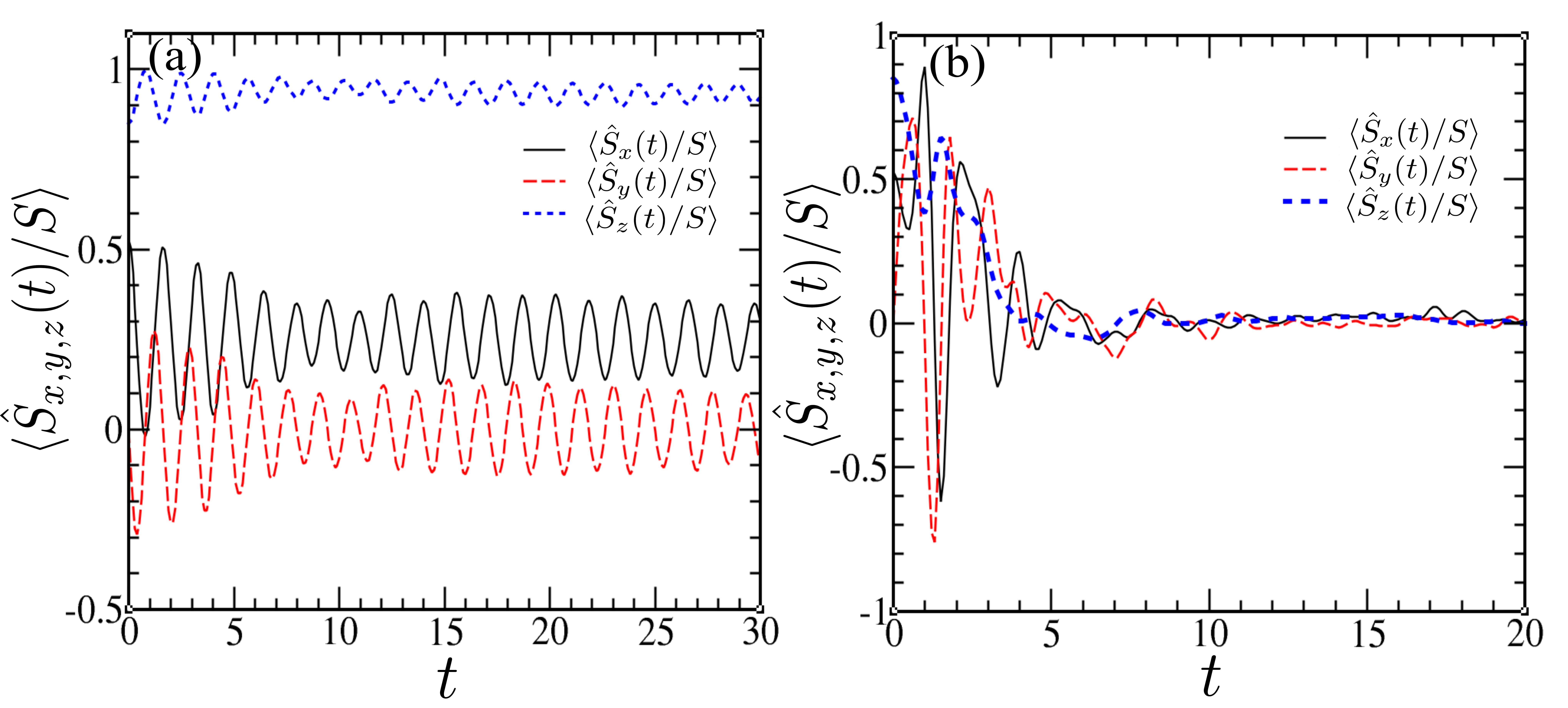}
    \caption{Dynamics of different initial states corresponding to non-ergodic vs ergodic regime. Time evolution of expectation values of observables (a) for initial state with $|z=0.85,\phi=0.0\rangle \otimes |q=-1.45,p=0.0\rangle$ having energy $E=-1.939<E_c$ and (b) for initial state with $|z=0.85,\phi=0.0\rangle \otimes |q=0.59,p=0.0\rangle$ having energy $E=1.506>E_c$. Parameters chosen are $S=30$, $\gamma/\gamma_c=3.0$, $U=0.5$ and $\omega_0=2.0$.}
    \label{fig:4}
\end{figure}

%%%%%%%%%%%%%%%%%%%%%%%%%%%%%%%%%%%%%%%%%%%%%%%%%%%%%%%%%%%%%%%%%%%%%%

%%%%%%%%%%%%%%%% NON ERGODIC VS ERGODIC STATES  %%%%%%%%%%%%%%%%%%%%
\begin{figure}[H]
\centering
\includegraphics[height=10cm,width=12.9cm]{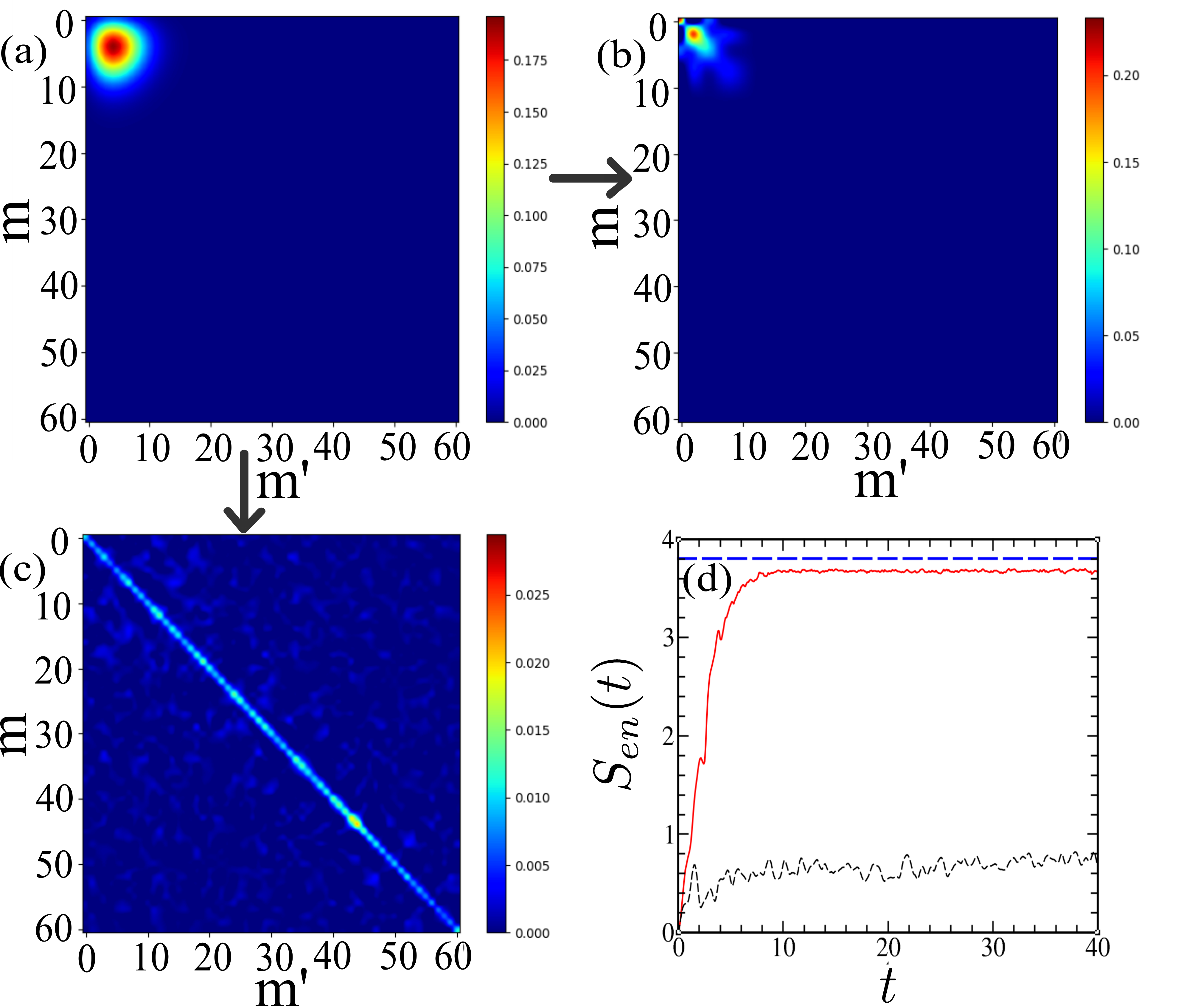}
\caption{a) shows $\hat{\rho}_S$ at t=0 with same initial condition in spin sector $\{z=0.85,\phi = 0.0\}$. (b) and (c) shows $\hat{\rho}_S$ at time $t=30$ for $|z=0.85,\phi=0.0\rangle \otimes |q=-1.45,p=0.0\rangle$ with $E=-1.939<E_c$ and for $|z=0.85,\phi=0.0\rangle \otimes |q=0.59,p=0.0\rangle$ with $E=1.506>E_c$ respectively. (d) shows time evolution of EE for $E<E_c$ (black dashed curve) and $E>E_c$ (red solid curve) where blue dashed line denotes $S_{max}$. Parameters chosen: $S=30$, $\gamma/\gamma_c=3.0$, $U=0.5$ and $\omega_0=2.0$. }
    \label{fig:5}
\end{figure}
%%%%%%%%%%%%%%%%%%%%%%%%%%%%%%%%%%%%%%%%%%%%%%%%%%%%%%%%%%%%%%%%%%%%%%

\noindent {\Large Phase Coherence:}

In BJJ, phase coherence between its two site is an important phenomena which signifies the macroscopic wave nature of the condensate. It is an interesting aspect to study the behaviour of phase coherence in non-equilibrium dynamics particularly in the ergodic regime. Quantum mechanically it can be calculated from the phase operator defining the relative phase between the two site \cite{Gati}. We can construct an orthonormal phase state basis consisting of $2S+1$ phase states with $\phi_m=\phi_0+2\pi m/(2S+1)$ where $m$ is a natural number $\in[0,2S]$ ,
\begin{eqnarray}
\ket{\phi_m}=\frac{1}{\sqrt{(2S+1)}}\sum^{S}_{n=-S}\exp(\imath n \phi_m)\ket{n}
\end{eqnarray}
%and $\phi_0$ is an arbitrary phase that is set to $-\pi$ for simplicity thus leading to $\phi\in[-\pi,\pi]$.
where $\phi_m\in[-\pi,\pi]$. The probability distribution is given by $p(\phi_m)=|\bra{\phi_m}\ket{\Psi}|^2$ with $\sum_{m}p(\phi_m)=1$. Then the phase coherence can be quantitatively measured by computing $\langle \cos{\phi} \rangle = \sum_{m}p(\phi_m)\cos{\phi_m}$. The decay of $\langle \cos{\phi} \rangle$ from unity to zero implies complete loss of phase coherence ($\langle\cos{\phi}\rangle=0$) from perfect coherence($\langle\cos{\phi}\rangle=1$). from our numerical study, we observe that phase coherence is retained during time evolution in the non-ergodic regime ($E<E_c$), while it is lost for initial state with $E>E_c$ which is expected due to ergodic evolution. This phenomena can be used to distinguish between the non-ergodic and ergodic regime experimentally.\\

\noindent {\Large Quantum Scars:}

As discussed above in the context of semiclassical dynamics, the steady state corresponding to the $\pi$-oscillation above QPT remains stable in the range $\gamma_c<\gamma<\gamma_u$. We study the imprint of $\pi$-oscillation from the Husimi distribution of the wavefunction at $E\approx E_0$ for increasing $\gamma$. In the stable region ($\gamma_c<\gamma<\gamma_u$), we observe that the probability amplitude of Husimi distribution is very localized at $\phi=\pm \pi$ (see Fig.\ref{fig:6}(a)) and above the instability point ($\gamma\geq\gamma_u$), we observe spreading of the probability distribution as seen from Fig.\ref{fig:6}(b). Significantly above the dynamical instability ($\gamma>\gamma_u$), interesting pattern in the Husimi distribution is observed as a reminiscence of dynamical orbits, as well the finite probability distribution around $\phi=\pm\pi$, which can be identified as scar of $\pi$-oscillation (see Fig.\ref{fig:6}(c)). 

%%%%%%%%%%%%%%%% SCARRED STATES AT E=1 %%%%%%%%%%%%%%%%%%%%
\begin{figure}[H]
    \centering
    \includegraphics[height=5.5cm,width=18cm]{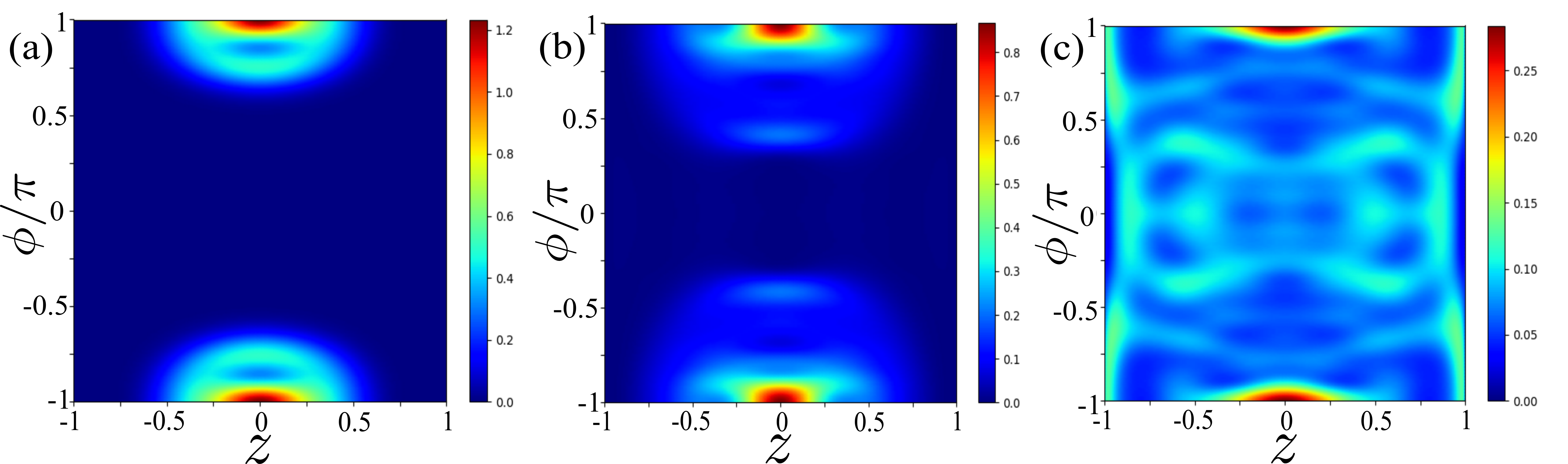}
    \caption{Husimi distribution of eigenstate at $E\approx E_0$ for (a) $\gamma/\gamma_c=1.25$, (b) $\gamma/\gamma_c=1.8$ and (c) $\gamma/\gamma_c=3.5$. Parameters chosen: $S=30$, $U=0.5$ and $\omega_0=3.0$. The unstable coupling strength $\gamma_u/\gamma_c=1.33$. }
    \label{fig:6}
\end{figure}
%%%%%%%%%%%%%%%%%%%%%%%%%%%%%%%%%%%%%%%%%%%%%%%%%%%%%%%%%%%%%%%%%%%%%%

\end{widetext}